\documentclass[10pt,a4paper,twocolumn,american,twocolumn,american,prb,a4paper]{revtex4}
\usepackage{lmodern}
\usepackage[T1]{fontenc}
\usepackage[latin9]{inputenc}
\usepackage{color}
\usepackage{array}
\usepackage{multirow}
\usepackage{amsmath}
\usepackage{amssymb}
\usepackage{graphicx}
\usepackage{wasysym}
\usepackage[final]{pdfpages}

\makeatletter

\@ifundefined{textcolor}{}
{%
 \definecolor{BLACK}{gray}{0}
 \definecolor{WHITE}{gray}{1}
 \definecolor{RED}{rgb}{1,0,0}
 \definecolor{GREEN}{rgb}{0,1,0}
 \definecolor{BLUE}{rgb}{0,0,1}
 \definecolor{CYAN}{cmyk}{1,0,0,0}
 \definecolor{MAGENTA}{cmyk}{0,1,0,0}
 \definecolor{YELLOW}{cmyk}{0,0,1,0}
}

\usepackage{slashed}

\makeatother

\usepackage{babel}
\begin{document}



# Erratum: "Quantitative analytical theory for disordered nodal points"

We point out an error in our previous arxiv submission 1704.08457. We discuss that the fRG flow equations cannot be integrated in a way that is mathematically rigorous. We present results of a second-order perturbation theory calculation.

In our previous arxiv submission 1704.08457, published as Phys. Rev. B 96, 064203(2017), we discussed how the functional renormalization group (fRG) can be used to study the effects of disorder in graphene and Weyl systems. Using a replica trick, we derived a set of fRG flow equations; their final form is given above Eq. (17) in the appendix. These flow equations are correct. In our arcticle, we solved them by simply integrating over the $\delta$-function $\delta(|i\omega| - \Lambda)$ that occurs on the rhs. This approach is not mathematically rigorous, which one can illustrate by considering the following, simplified case.

If in the flow equation for the self-energy $\Sigma_\Lambda$ we replace $A$ by the bare vertex $A_\infty$ and also assume uncorrelated disorder (in which case $\Lambda$ serves as UV-cutoff), the self-energy takes the form $\Sigma_\Lambda(i\eta, k) = -i\Gamma_\Lambda$ with $\Gamma_\infty = 0$. For $\hbar v \Lambda \gg \Gamma_\Lambda, \eta$, its flow equation is given by

$$\partial_\Lambda \Gamma_\Lambda = -\delta(\eta - \Lambda) \frac{K}{2\pi} (\Gamma_\Lambda + \eta) \log\left(\frac{\hbar v \Lambda}{\Gamma_\Lambda + \eta}\right). \quad (1)$$

This equation can be understood as an impulsive differential equation of the form $\partial_\Lambda f_\Lambda = \delta(\Lambda) g(f_\Lambda)$, which is analyzed in the mathematical literature. A rigorous solution to such a type of equation was constructed, e.g., in Ref. [1]. This construction, however, requires the function $g$ to be Lipschitz, which is not the case in Eq. (1) in the limit $\eta \to 0$. Thus, the theory of Ref. [1] does not apply. Despite considerable efforts, we have not succeeded in finding a mathematically rigorous solution to our flow equations (the equations above (17) in the appendix), which are not Lipschitz. This problem can be cured by using different fermionic fRG schemes, e.g., a Wick-ordered scheme with a k-space cutoff [2] or a 2PI scheme [3]. Alternatively, one could employ a partially bosonized formulation where the bosonic vacuum expectation value relates to the fermionic self-energy. Results on this will be presented elsewhere.

The fact that our solution to the flow equations is flawed can also be seen by discussing the relationship with perturbation theory. In particular, the effects of static disorder on the electronic Green function can be expressed as an infinite perturbation expansion for the electronic self-energy. The two terms of second order in disorder strength $K$ are "crossed" and "rainbow" diagrams similar as in Fig. 3 (ii.a) and (ii.b), respectively, but with bare Green functions (single fermion lines). Our self-consistency equation in Fig. 3 (with dressed fermion lines) was derived from the integration of (truncated) fRG flow equations. Expanding to second order in $K$ yields the aforementioned "crossed" and "rainbow" diagrams, but from diagram (i), another rainbow diagram is obtained by expanding the dressed fermion propagator. Thus, our fRG solution suffers from double-counting, which should not occur to the order of the truncation.

To summarize: Eq. (17) and the ensuing diagrammatic representation shown in Fig. 3 cannot be the proper solution to the preceeding, correct set of fRG flow equations, since the $\delta$-function cannot be naively integrated over. We have found no way to construct a mathematically rigorous solution; cures using different fRG schemes will be presented elsewhere. However, if we simply exclude diagram (ii.b) in Fig. 3, the resulting equation represents second order self-consistent perturbation theory (SCPT2). In Fig. E1, we reproduce the bottom panels of Figs. 1 and 2 but replace the fRG data by the results of SCPT2 (purple line) which improves on the first order self-consistent perturbation theory (SCBA, blue line).

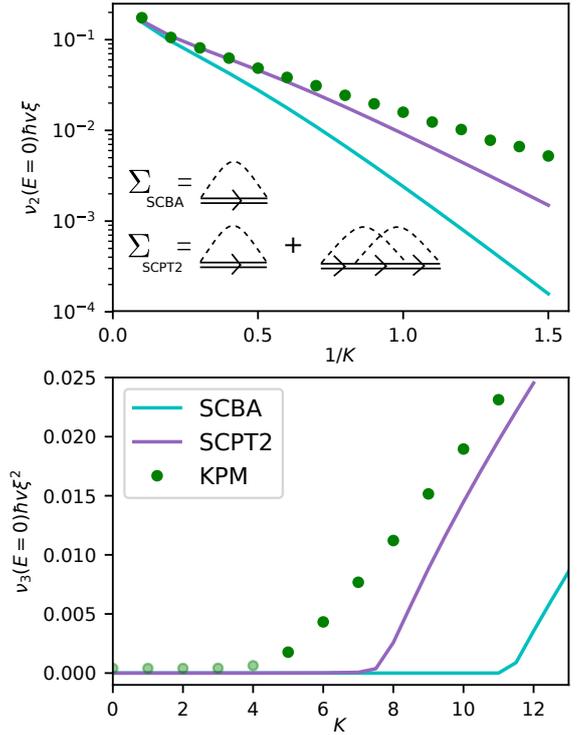

Figure E1: Zero energy density of states of a two-dimensional disordered Dirac node [$\nu_2(E = 0)$, top] and of a three-dimensional disordered Weyl node [$\nu_3(E = 0)$, bottom] as calculated by the Kernel Polynomial method (KPM) versus disorder strength $K$ (green dots). The exact KPM data is compared to the SCBA (blue line) and self consistent perturbation theory of 2nd order (SCPT2, purple line).



\title{Quantitative analytical theory for disordered nodal points}

\author{Björn Sbierski, Kevin A. Madsen, Piet W. Brouwer, Christoph Karrasch}

\affiliation{Dahlem Center for Complex Quantum Systems and Institut für Theoretische
Physik, Freie Universität Berlin, D-14195, Berlin, Germany}
\begin{abstract}
Disorder effects are especially pronounced around nodal points in
linearly dispersing bandstructures as present in graphene or Weyl
semimetals. Despite the enormous experimental and numerical progress,
even a simple quantity like the average density of states cannot be
assessed quantitatively by analytical means. We demonstrate how this
important problem can be solved employing the functional renormalization
group method and, for the two dimensional case, demonstrate excellent
agreement with reference data from numerical simulations based on
tight-binding models. In three dimensions our analytic results also
improve drastically on existing approaches.
\end{abstract}

\date{\today}

\maketitle
\section{Introduction} 
Two dimensional graphene \cite{Novoselov2005}
and three dimensional Weyl materials \cite{Bernevig2015} are important
examples of Dirac type semimetals. Their electronic structure features
a nodal degeneracy point where two linearly dispersing Bloch bands
meet. Due to the vanishing density of states (DOS), disorder effects
can be expected to be particularly pronounced in these materials and
have been actively studied, for reviews see Refs. \cite{DasSarma2011,Syzranov2016c}.
Despite all this effort on the disorder problem for nodal points,
analytical results, even for a quantity as simple as the DOS, are
at best qualitatively correct but fail widely in their quantitative
predictions, even for weak disorder. This is surprising insofar as
exact answers can be obtained with ease from numerical simulations
of non-interacting lattice Hamiltonians. The scope of this work is
to show how tremendous progress on this long-standing problem can
be achieved by employing a variant of the functional renormalization
group (fRG).

We consider the minimal continuum model of a single disordered node
in $d\!=\!2,3$ dimensions, 
\begin{equation}
H_{d}=H_{0,d}+U_{d},\label{eq:Hd}
\end{equation}
where $H_{0,2}=\hbar v(\sigma_{x}k_{x}+\sigma_{y}k_{y})$ is a $d\!=\!2$
Dirac Hamiltonian and $H_{0,3}=\hbar v(\sigma_{x}k_{x}+\sigma_{y}k_{y}+\sigma_{z}k_{z})$
a $d\!=\!3$ Weyl Hamiltonian written with the standard Pauli matrices
$\sigma_{i=x,y,z}$. The disorder potential $U_{d}(\mathbf{r})$,
taken to be proportional to the unit matrix, is commonly assumed to
have Gaussian correlations and zero mean. Explicitly, we assume a
smooth form of the correlator
\begin{equation}
\mathcal{K}_{d}\left(\mathbf{r}-\mathbf{r}^{\prime}\right)=\left\langle U_{d}(\mathbf{r})U_{d}(\mathbf{r}^{\prime})\right\rangle =K\frac{(\hbar v)^{2}}{(2\pi)^{d/2}\xi^{2}}e^{-|\mathbf{r}-\mathbf{r}^{\prime}|^{2}/2\xi^{2}},\label{eq:smooth correlator}
\end{equation}
where $\left\langle ...\right\rangle $ denotes the disorder average.
As $H_{0,d}$ is lacking any scale, the disorder correlation length
$\xi$ serves as the fundamental scale in the problem. The dimensionless
parameter $K$ measures the disorder strength. In the Brillouin zone
of real materials, nodal points usually come in pairs. This is enforced
by symmetry (graphene) or topology (Weyl). However, these pairs can
have a sizable k-space separation $\Delta k$. If $\xi\Delta k\gg1$
the intra-node scattering dominates over inter-node scattering and
the model (\ref{eq:Hd}) is a reasonable low-energy approximation for realistic
materials.

While Eq. (\ref{eq:Hd}) with the correlator (\ref{eq:smooth correlator})
has the advantage that it can be easily approximated in tight-binding
models if $\xi\gg a$ ($a$ being the lattice scale) another common
choice for $\mathcal{K}_{d}$ more convenient for analytical calculations
is the white noise limit $\xi\rightarrow0$, 
\begin{equation}
\mathcal{K}_{d}^{GWN}\left(\mathbf{r}\right)=K(\hbar v)^{2}\xi^{d-2}\delta\left(\mathbf{r}\right),\label{eq:GWN correlator}
\end{equation}
along with the prescription that $1/\xi$ serves as an ultraviolet cutoff
for the clean dispersion $H_{0,d}$. We will use the white noise approximation
to make contact with known results.

The bulk DOS can be calculated as 
\begin{equation}
\nu\left(E\right)=-\frac{1}{\pi}\mathrm{Im}\,\mathrm{Tr}\int_{\mathbf{k}}G_{\mathbf{k}}^{R}\left(E\right),\label{eq:DOS}
\end{equation}
where $\int_{\mathbf{k}}=\left(2\pi\right)^{-d}\int d\mathbf{k}$
 and $G_{\mathbf{k}}^{R}\left(E\right)$ is the retarded (matrix-valued)
Green function. For the clean Hamiltonian $H_{0,d}$, one has $\nu_{0,d}\left(E\right)=|E|^{d-1}/(2\pi)^{d-1}\left(\hbar v\right)^{d}$,
vanishing at the degeneracy point. If disorder is thought of as
a local chemical potential creating carriers from conduction or valence
bands, a finite $\nu_{d}\left(E\!=\!0\right)$ can be expected (since disorder
is a self-averaging quantity, we omit $\left\langle ...\right\rangle $).
In the following, we distinguish between 'numerical' approaches based
on explicit generation of a large number of random disorder realizations
$U_{d}(\mathbf{r})$ in Eq. (\ref{eq:Hd}) and 'analytical' methods
starting from Eq. (\ref{eq:smooth correlator}). While the former are
well established, up to now there is no known analytical method that
could reproduce numerical results with reasonable accuracy, not even
for small $K$.

The scope of this work is to show how this long-standing problem can
be solved by a variant of the functional renormalization group (fRG)
which allows to rewrite the disorder problem as an \textemdash{} in
principle infinite \textemdash{} hierarchy of coupled self-consistency
equations for vertex functions. We apply this technique to calculate
the DOS $\nu_{d}\left(E=0\right)$ and find that even a simple truncation
of the above hierarchy yields results in very good quantitative agreement
with numerically exact data obtained from the kernel polynomial method
at much higher computational costs. We acknowledge an earlier study
by Katanin \cite{Katanin2013} with similar objectives but a different
variant of the fRG. However, our results go significantly beyond those
of Ref. \cite{Katanin2013}, where only $d\!=\!2$ was investigated without
comparison to numerically exact results. 

\section{Exact numerical DOS} To gauge the quality of analytical
approaches discussed in the remaining sections, let us start by obtaining
numerically exact DOS data for the Dirac and Weyl systems with smooth
disorder, described by Eqns. (\ref{eq:Hd}) and (\ref{eq:smooth correlator}).
We apply the kernel polynomial method (KPM) \cite{Weisse2006}, a
numerically efficient tool to approximate the DOS of large lattice
Hamiltonians $H$ represented as sparse matrices. The DOS $\nu\left(E\right)$
as a function of energy $E$ is expanded in Chebyshev polynomials
and the expansion coefficients $\mu^{(n)}$ are expressed as a trace
over a polynomial in $H$. Using recursion properties of Chebyshev
polynomials, the $\mu^{(n)}$ can be efficiently computed (up to order $N$) involving only sparse matrix-vector products and
a statistical evaluation of the trace.

\begin{figure}
\noindent \begin{centering}
\includegraphics{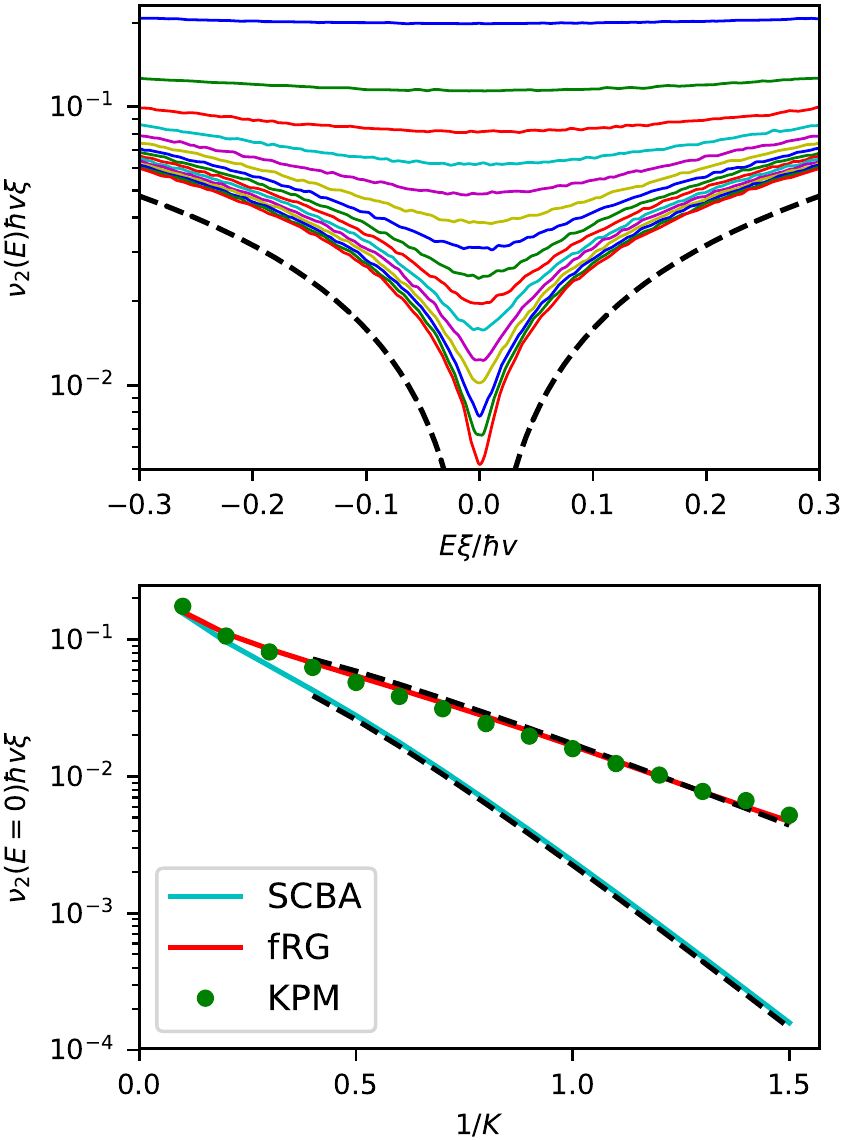}
\par\end{centering}
\caption{\label{fig:DOS_2dDirac}{\small{}Top: Density of states $\nu_{2}$
for a two-dimensional disordered Dirac node as a function of energy
$E$ as calculated by the KPM for various disorder strengths $K$
(for values of $K$ c.f. bottom panel). The dashed line denotes the
analytic result for the clean case. Bottom: The zero energy density
of states $\nu_{2}(E=0)$ from KPM (dots) compared to the SCBA (blue
line) and fRG (red line). The parameters for the simulation are $\xi=3a$
(except for the two largest $K$, where $\xi=4a$), linear system
size $L=2000\xi$, $20$ random vectors for calculating the trace
and an expansion order of up to $15000$ moments. The data represents
an average over $20$ disorder realizations and is normalized to a
single node. The dashed lines denote fits to the white noise forms of the density of states from SCBA and RG as discussed in the main text.}}
\end{figure}

\begin{figure}
\noindent \begin{centering}
\includegraphics{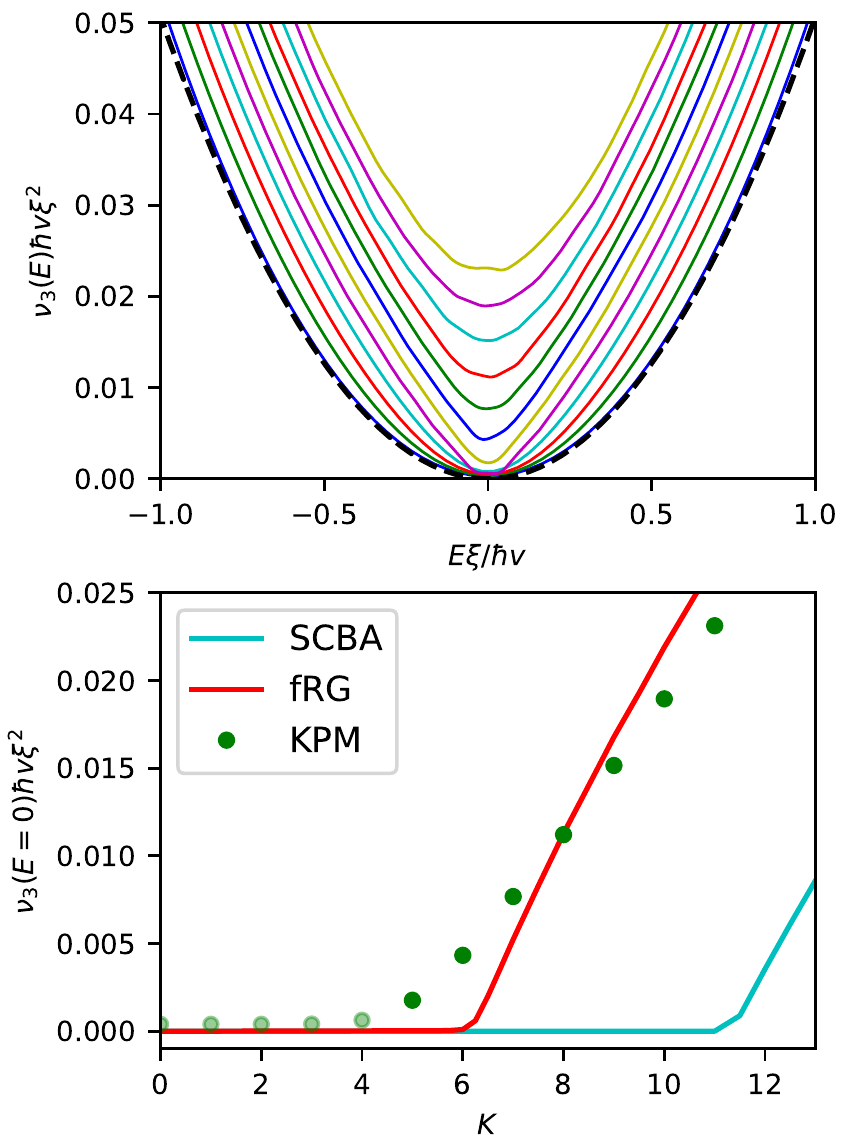}
\par\end{centering}
\caption{\label{fig:DOS_3dWeyl}{\small{}Top: Density of states $\nu_{3}$
for a three-dimensional disordered Weyl node as a function of energy
$E$ as calculated by the KPM for disorder strengths $K=0,1,2,...,11$
(bottom to top). The dashed line denotes the analytic result for the
clean case. Bottom: The zero energy density of states $\nu_{3}(E=0)$
from KPM (dots) compared to the SCBA (blue line) and fRG (red line).
The parameters for the simulation are $\xi=4a$, linear system size
$L=180\xi$, $20$ random vectors for calculating the trace and an
expansion order of up to 2000 moments. The data represents an average
over $40$ disorder realizations and is normalized to a single node.
The semi-transparent data points for $K\leq4$ suffer from finite-$L$
effects and overestimate the true bulk DOS.}}
\end{figure}

The clean nodal Hamiltonian $H_{0,d}$ is approximated as the low
energy theory of the following tight-binding models on a square/cubic
lattice (with constant $a$, size $L^{d}$)
\begin{align}
H_{0,d}^{\mathrm{L}} & =\!\frac{\hbar v}{a}\!\begin{cases}
\sigma_{x}\cos ak_{x}\!+\!\sigma_{y}\cos ak_{y} & \!\!\!\!(d\!=\!2)\\
\sigma_{x}\sin ak_{x}\!+\!\sigma_{y}\sin ak_{y}\!-\!\sigma_{z}\cos ak_{z} & \!\!\!\!(d\!=\!3),
\end{cases}\label{eq:H_lattice}
\end{align}
which feature four/eight nodal points for $d\!=\!2$ and $d\!=\!3$,
respectively, with minimal mutual distance $\Delta k=\frac{\pi}{a}$.
We apply periodic boundary conditions and add a correlated disorder
potential as in Eq. (\ref{eq:smooth correlator}). If our disordered
lattice model would faithfully emulate the continuum Hamiltonian (\ref{eq:Hd}),
the DOS at zero energy must be of the scaling form $\nu_{d}\left(E=0\right)=\left(\hbar v\right)^{-1}\xi^{1-d}f\left(K\right)$
with $f(K)$ a dimensionless function. We have checked that the KPM
data based on the lattice Hamiltonian Eq. (\ref{eq:H_lattice}) fulfills
this scaling condition once $\xi\gg a$ so that (i) the smooth disorder
correlations are well represented on the discrete lattice, (ii) the
disorder induced energy scale is well below the scale of order $\hbar v/a$
where $H_{0,d}^{\mathrm{L}}$ deviates from $H_{0,d}$ and (iii) the
inter-node scattering rate is sufficiently suppressed compared to
the intra-node rate (the factor is $\exp[-(\Delta k)^{2}\xi^{2}/2]$).
Moreover, we require $L\gg\xi$ to suppress finite-size effects. Thus,
the KPM data (normalized to a single node) shown as dots in Fig. \ref{fig:DOS_2dDirac}
($d=2$) and Fig. \ref{fig:DOS_3dWeyl} ($d=3$) can be regarded as
the exact zero energy DOS of the continuum model Eq. (\ref{eq:Hd}).
Simulation parameters are given in the figure captions. In spite of
the abundant literature on similar numerical studies for the DOS of
disordered 2d Dirac (see Refs. \cite{Peres2006,Pereira2008,Wu2008,Ziegler2009a})
and 3d Weyl systems (see Refs. \cite{Kobayashi2014,Pixley2015a,Trescher2016}),
we are not aware of existing high-precision data obtained for a smooth
disorder correlator and with the required scaling properties fulfilled.

\section{Disordered $d\!=\!2$ Dirac node} We proceed by
discussing existing analytical approaches to the disorder problem
in the $d\!=\!2$ Dirac case. The self-consistent Born approximation (SCBA)
determines the disorder induced self-energy $\Sigma\equiv G^{-1}-G_{0}^{-1}$
(where $G_{0}$ is the Green function of the clean system) according
to the diagram in Fig. \ref{fig:Diagrams}(i) \cite{Shon1998,Ostrovsky2006,Noro2010a}.
The corresponding self-consistent equation can be solved in closed
form for the white noise correlator Eq. (\ref{eq:GWN correlator})
and yields a disorder induced scale $\Gamma=\frac{\hbar v}{\xi}e^{-2\pi/K}$
(for $K\lesssim1$) exponentially small in $K$ appearing in the imaginary
self-energy $\Sigma=\pm i\Gamma$ and a DOS $\nu_{2}\left(E=0\right)\hbar v\xi\propto e^{-2\pi/K}/K$
\cite{Ostrovsky2006}. In Fig. \ref{fig:DOS_2dDirac} (bottom panel),
this result (dashed line) compares well to the DOS obtained from the
SCBA with smooth disorder correlator (\ref{eq:smooth correlator})
(blue line). However, comparing to the exact KPM-DOS
in Fig. \ref{fig:DOS_2dDirac} (dots), we find that albeit the exponential
form is correctly predicted by the SCBA, the slope (prefactor in the
exponent) is roughly a factor 2 off. 

The failure of the SCBA can be attributed to interference corrections
from multiple disorder scattering events \cite{Aleiner2006}, see
diagrams (ii.a) and (ii.b) for the lowest order corrections. While
unimportant in ordinary metals (where $1/k_{F}l\ll1$ with $k_{F}$
Fermi wavevector and $l$ the mean free path serves as a small parameter),
for Dirac materials these diagrams provide corrections of order $\ln\left[\hbar v/\xi\Gamma\right]$.
Accordingly, their contribution vanishes for strong disorder where
the SCBA becomes reliable, c.f. Fig. \ref{fig:DOS_2dDirac}.

To go beyond the SCBA, Refs. \cite{Ostrovsky2006,Aleiner2006} used
the super-symmetry method. Alternatively, the replica trick \cite{Altland2006}
can be employed: It takes a disorder average over $R$ copies (replicas)
of the original problem seeing the same disorder potential. The resulting
action $S=S_{d,0}+S_{d,dis}$ is translational invariant but contains,
besides the free part $S_{d,0}=\sum_{\alpha=1}^{R}\int_{\omega}\int_{\mathbf{k}}\sum_{\sigma,\sigma^{\prime}}\bar{\psi}_{\omega\mathbf{k}\sigma^{\prime}}^{\alpha}(i\omega-H_{0,d})_{\sigma^{\prime}\sigma}\psi_{\omega\mathbf{k}\sigma}^{\alpha}$
an attractive inter-replica interaction which is elastic (i.e. without
frequency transfer)
\begin{align}
S_{d,\mathrm{dis}} & =\sum_{\alpha,\beta=1}^{R}\int_{\omega_{1},\omega_{2}}\int_{\mathbf{k}_{1}^{\prime},\mathbf{k}_{1},\mathbf{k}_{2}^{\prime},\mathbf{k}_{2}}2\pi\delta_{\mathbf{k}_{1}^{\prime}-\mathbf{k}_{1}+\mathbf{k}_{2}^{\prime}-\mathbf{k}_{2}}\label{Sdis}\\
\times & \frac{-\mathcal{K}_{d}\left(\mathbf{k}_{1}^{\prime}-\mathbf{k}_{1}\right)}{2}\sum_{\sigma,\sigma^{\prime}}\bar{\psi}_{\omega_{1}\mathbf{k}_{1}^{\prime}\sigma}^{\alpha}\psi_{\omega_{1}\mathbf{k}_{1}\sigma}^{\alpha}\bar{\psi}_{\omega_{2}\mathbf{k}_{2}^{\prime}\sigma^{\prime}}^{\beta}\psi_{\omega_{2}\mathbf{k}_{2}\sigma^{\prime}}^{\beta}.\nonumber 
\end{align}
Assuming the white noise correlator (\ref{eq:GWN correlator}) that
comes with the UV cutoff $1/\xi$ in k-space, this action is susceptible
to a Wilsonian momentum-shell RG analysis \cite{Ostrovsky2006,Aleiner2006,Schuessler2009}.
Successively integrating out high energy modes down to $\lambda^{-1}/\xi$
($\lambda\geq1$) perturbatively, the action can be approximately
mapped to itself with rescaled momenta, fields and coupling constants.
If the velocity is kept constant, the two-loop RG equation for the
flowing disorder strength $\tilde{K}\left(\lambda\right)$ reads \cite{Schuessler2009}
\begin{equation}
d\tilde{K}/d\,\mathrm{ln}\lambda=\tilde{K}^{2}/\pi+\tilde{K}^{3}/(2\pi^{2}).\label{eq:Ostrovsky flow}
\end{equation}
Starting with the initial condition $\tilde{K}\left(1\right)\!=\!K$
the flow is to strong coupling where the perturbation theory leading
to Eq. (\ref{eq:Ostrovsky flow}) breaks down. To find the energy
scale $\Gamma$ where this happens (and below which the DOS is presumably
constant), let us assert $\tilde{K}\left(\hbar v/\Gamma\xi\right)\sim1$
which, in the limit of $K\ll1$, leads to $\Gamma\propto\frac{\hbar v}{\xi}\sqrt{\frac{1}{K}}e^{-\pi/K}$
\cite{Schuessler2009} correcting for the factor 2 in the exponent
as found from the SCBA. The DOS at the nodal point is expected to
be governed by this emergent energy scale $\nu_{2}\left(E=0\right)\hbar v\xi\propto\Gamma$,
in agreement with the KPM results in Fig. \ref{fig:DOS_2dDirac}.

The Wilsonian RG calculation gave the correct exponential scale governing
the disorder problem. However, it is not quantitative in the sense
that numerical estimates for, say, the DOS could be obtained in the
strong coupling limit. We will now show how the fRG method overcomes
the difficulties mentioned above and use it to obtain quantitative
results for the disorder induced DOS at the nodal point without any
fitting parameters. 

\section{fRG approach} The fRG \cite{Metzner2012} introduces
a flow parameter $\Lambda$ in the bare propagator and rewrites the
many-body problem in a hierarchy of coupled flow equations for vertex
functions with respect to $\Lambda$. The flow parameter is chosen
such that for $\Lambda=\infty$, the vertex functions are known exactly
and for $\Lambda=0$ the original problem is retained. We relegate
a detailed discussion of technicalities to the appendix and only
highlight the most important points and modifications related to use
of the fRG with the replicated action. 

\begin{figure}
\noindent \begin{centering}
\includegraphics{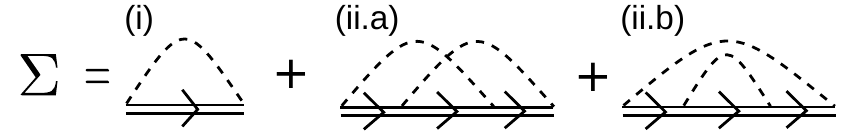}
\par\end{centering}
\caption{\label{fig:Diagrams}{\small{}Diagrammatic representation of the self-consistency
equation (\ref{eq:self-consistency_short}) for the disorder induced
self-energy $\Sigma$ as obtained from the $\mathcal{O}(K^{2})$ truncation
of the fRG. Diagram (i) is the first order term equivalent to SCBA
while (ii.a) and (ii.b) are of second order in $K$. Dashed lines
denote disorder correlators and double lines self-energy dressed Green
functions.}}
\end{figure}

To actually calculate expectation values and vertex functions from
the replicated action, the replica limit $\left\langle O\right\rangle =\underset{R\rightarrow0}{\mathrm{lim}}\frac{1}{R}\sum_{\alpha=1}^{R}\left\langle \mathcal{O}\left(\bar{\psi}^{\alpha},\psi^{\alpha}\right)\right\rangle _{\psi}$
is required, where $\left\langle \mathcal{O}\left(\bar{\psi},\psi\right)\right\rangle _{\psi}=\int D(\bar{\psi},\psi)\mathcal{O}\left(\bar{\psi},\psi\right)e^{-S\left[\bar{\psi},\psi\right]}$
stands for the standard functional average over a polynomial of fields
$\mathcal{O}\left(\bar{\psi},\psi\right)$ \cite{Altland2006}. In
a peturbative expansion (which is also at the heart of the fRG flow
equations), thus only diagrams without closed fermion loops have a
finite contribution in the replica limit. This also means that mixing
of replica indices in the relevant diagrams is avoided. One can also
show that the elastic nature of the interaction vertex derived from
(\ref{Sdis}) is maintained along the flow. As a consequence, on the
right hand side (rhs) of the flow equations the frequency integral
as required for inelastic (true) interactions, is absent. Thus introducing
$\Lambda$ via a Matsubara frequency cutoff scheme results in a Dirac
delta function on the rhs which allows for a direct integration of the corresponding
flow equations and results in a self-consistent hierarchy of equations
for the vertices. So far no approximations have been made. To proceed,
we truncate the hierarchy to order $K^{2}$. This is a pragmatic choice,
that still goes beyond all diagrammatic schemes previously applied
to disordered Dirac materials explicitly. Subsequently, we eliminate
the interaction vertex in favor of the self-energy. The remaining
self-consistency equation reads
\begin{align}
\Sigma\left(\mathbf{k}\right) & =K\left(\hbar v\right)^{2}\int_{\mathbf{q}}G(\mathbf{q})e^{-\frac{1}{2}\xi^{2}\left|\mathbf{q}-\mathbf{k}\right|^{2}}\label{eq:self-consistency_short}\\
 & +K^{2}\left(\hbar v\right)^{4}\int_{\mathbf{q},\mathbf{p}}e^{-\frac{1}{2}\xi^{2}(\left|\mathbf{k}-\mathbf{p}\right|^{2}+\left|\mathbf{q}-\mathbf{p}\right|^{2})}\nonumber \\
 & \times G(\mathbf{p})\cdot G(\mathbf{q})\cdot\left[G(\mathbf{k}+\mathbf{q}-\mathbf{p})+G(\mathbf{p})\right],\nonumber 
\end{align}
and is displayed in Fig. \ref{fig:Diagrams} diagrammatically: The
term of order $K$ represents the SCBA approximation, c.f. diagram
(i), the two second order terms are shown in diagrams (ii.a) and (ii.b)
respectively. Although these diagrams would also appear in perturbation
theory, the fRG approach (i) rigorously justifies the use of the self-energy
dressed propagators and (ii) indicates how we could consistently go
beyond order $K^{2}$ by allowing feedback for the vertex self-consistency
equation. 

To solve Eq. (\ref{eq:self-consistency_short}), we parameterize the
self-energy using polar ($d\!=\!2$) or spherical ($d\!=\!3$) coordinates
and proceed by iteration. We compute the DOS from Eq. (\ref{eq:DOS}).
Further details are given in the appendix. In the $d=2$ Dirac case,
the resulting DOS (red line) shows excellent agreement with the numerically
exact KPM data and justifies the used order $K^{2}$ truncation a
posteriori, well capable of capturing the exponential scale derived
from Eq. (\ref{eq:Ostrovsky flow}).

On the pragmatic side, let us note that our fRG method also has advantages
over the KPM method besides being analytic. For example, in Fig. \ref{fig:DOS_2dDirac},
the KPM data for $\nu_{2}\left(E\right)$ shows a dip around $E=0$
that can only be resolved for small $K$ if the system size $L$ and
expansion order $N$ is taken large. In comparison, the solution of
Eq. (\ref{eq:self-consistency_short}) requires only a small fraction
of computational effort.

\section{Disordered $d\!=\!3$ Weyl node} We now turn to
the disorder induced DOS for a $d=3$ Weyl node. Here, weak disorder
is irrelevant so that the DOS is maintained at zero. Only for $K>K_{c}$,
disorder induces a finite DOS, see Fig. \ref{fig:DOS_3dWeyl} for
the KPM data (dots). These qualitative features were correctly predicted
by the SCBA (blue line, see Refs. \cite{Fradkin1986,Fradkin1986a,Biswas2014,Ominato2015})
and by the momentum shell RG treatment, see Refs. \cite{Syzranov2014,Syzranov2016}.
From the KPM, we find $K_{c}^{KPM}=4\pm0.5$ (the precision is limited
by finite size effects) while $K_{c}^{SCBA}\simeq11$ (blue line)
is off by more than a factor two. The one-loop RG result $K_{c}^{RG_{1}}=\pi^{2}\simeq10$
can be improved with respect to the KPM value by adding two-loop corrections
$K_{c}^{RG_{2}}=\pi^{2}/2\simeq5$. However, quantitative predictions
for the DOS in the strong-disorder phase cannot be obtained with the
RG approach.

When compared to the $d\!=\!2$ case, the additional challenge for
the fRG approach in the Weyl case is that the interesting disorder
strengths $K\apprge K_{c}$ are not numerically small. Thus we assume
that our $\mathcal{O}(K^{2})$ truncation of the fRG equations might
cause a sizable error. Surprisingly, the fRG results (red line) yield
$K_{c}^{fRG}\simeq6$ and predict the available exact DOS for $K>7$
within an error of a few percent. On the one hand, we expect that
the remaining numerical error of the fRG method could be systematically
reduced by considering the fRG flow of the interaction vertex, which
we leave for future research. On the other hand, this might not improve
the accuracy for $K\simeq K_{c}$ where rare region effects which
lie beyond any order of perturbation theory, are expected to dominate
the DOS \cite{Nandkishore2014,Pixley2016,Pixley2017,Holder2017}. However, it
is known that their influence can be suppressed by choosing a different
disorder model \cite{Pixley2016a}.

\section{Conclusion} We applied the fRG to treat the disorder
problem at nodal points in two and three dimensions. From the resulting
hierarchy of self-consistency equations, we calculate the bulk DOS
and show that it is superior in accuracy to any other existing analytical
approach. Suprisingly, for two dimensions, a truncation of the self-consistency equation at second order of $K$ is sufficient, while in three dimensions the accuracy could probably be increased with increasing order. We leave this suggestion for future work, along with the calculation of
other experimentally relevant transport properties from fRG. More complicated
disorder models, in particular vector disorder in two dimensions and its characteristic $\nu(E)$ behavior or scattering between multiple nodal points, as present
in realistic materials, could also be studied in the future.

\section*{Acknowledgments}We thank Pavel Ostrovsky and Voker Meden for
useful discussions. Numerical computations were done on the HPC cluster
of Fachbereich Physik at FU Berlin. Financial support was granted
by the Deutsche Forschungsgemeinschaft through the Emmy Noether program
(KA 3360/2-1) and the CRC/Transregio 183 (Project A02).\emph{ }

\bibliographystyle{apsrev4-1}
\bibliography{library}

\clearpage{}

\onecolumngrid

\section*{Appendix: fRG with replicated action and solution of self-consistency
equation}

\paragraph{fRG flow equations and vertex structure for the replica interaction.\textemdash{}}

The fRG flow equations for the self energy $\Sigma$ and the interaction
vertex $\Gamma$ have the form \cite{Metzner2012}
\begin{equation}
\partial_{\Lambda}\Sigma_{\Lambda}\left(1^{\prime};1\right)=-\int_{2,2^{\prime}}\left[\dot{G}_{\Lambda}\right]_{2,2^{\prime}}\Gamma_{\Lambda}\left(1^{\prime},2^{\prime};1,2\right)\label{eq:D_Sigma_general}
\end{equation}
and, in three-particle vertex truncation,
\begin{eqnarray}
\partial_{\Lambda}\Gamma_{\Lambda}\left(1^{\prime},2^{\prime};1,2\right) & = & \int_{3,3^{\prime},4,4^{\prime}}\label{eq:D_Gamma_general}\\
 &  & \Gamma_{\Lambda}\left(1^{\prime},2^{\prime};3,4\right)\left(\left[G_{\Lambda}\right]_{3,3^{\prime}}\left[\dot{G}_{\Lambda}\right]_{4,4^{\prime}}\right)\Gamma_{\Lambda}\left(3^{\prime},4^{\prime};1,2\right)\nonumber \\
 & + & \Gamma_{\Lambda}\left(1^{\prime},4^{\prime};3,2\right)\left(\left[G_{\Lambda}\right]_{3,3^{\prime}}\left[\dot{G}_{\Lambda}\right]_{4,4^{\prime}}+\left[G_{\Lambda}\right]_{4,4^{\prime}}\left[\dot{G}_{\Lambda}\right]_{3,3^{\prime}}\right)\Gamma_{\Lambda}\left(2^{\prime},3^{\prime};4,1\right)\nonumber \\
 & - & \Gamma_{\Lambda}\left(1^{\prime},4^{\prime};3,1\right)\left(\left[G_{\Lambda}\right]_{3,3^{\prime}}\left[\dot{G}_{\Lambda}\right]_{4,4^{\prime}}+\left[G_{\Lambda}\right]_{4,4^{\prime}}\left[\dot{G}_{\Lambda}\right]_{3,3^{\prime}}\right)\Gamma_{\Lambda}\left(2^{\prime},3^{\prime};4,2\right),\nonumber 
\end{eqnarray}
where $\dot{G}_{\Lambda}=G_{\Lambda}(\partial_{\Lambda}[G_{0,\Lambda}^{-1}])G_{\Lambda}$
is the single-scale propagator and the multi-index $\left\{ \alpha_{1}i\omega_{1}\mathbf{k}_{1}\sigma_{1}\right\} \equiv1$
includes the relevant single-particle indices: replica index, Matsubara
frequency, momentum and spin, respectively. We also use the notation
$1_{\alpha_{j}}\equiv\left\{ \alpha_{j}i\omega_{1}\mathbf{k}_{1}\sigma_{1}\right\} $
and $1_{\alpha_{j}i\omega_{k}}\equiv\left\{ \alpha_{j}i\omega_{k}\mathbf{k}_{1}\sigma_{1}\right\} $
at our convenience and also abbreviate integrals and sums on the rhs
as $\int_{1}\equiv\sum_{\alpha_{1}}\,\frac{1}{2\pi}\int d\omega_{1}\,\frac{1}{(2\pi)^{d}}\int\mathrm{d}\mathbf{k}_{1}\,\sum_{\sigma_{1}}$.

Starting from the inter-replica interaction $S_{\mathrm{dis}}$, Eq.
(\ref{Sdis}) in the main text, we find the bare vertex by anti-symmetrization
\begin{eqnarray}
\Gamma_{\infty}\left(1^{\prime},2^{\prime};1,2\right) & = & 2\pi\delta_{i\omega_{1}^{\prime}-i\omega_{1}}\delta_{\alpha_{1}^{\prime},\alpha_{1}}\,2\pi\delta_{i\omega_{2}^{\prime}-i\omega_{2}}\delta_{\alpha_{2}^{\prime},\alpha_{2}}\,A_{\infty}\left(\mathbf{k}_{1}^{\prime}\sigma_{1}^{\prime},\mathbf{k}_{2}^{\prime}\sigma_{2}^{\prime};\mathbf{k}_{1}\sigma_{1},\mathbf{k}_{2}\sigma_{2}\right)\label{eq:bare-vertex}\\
 & - & 2\pi\delta_{i\omega_{2}^{\prime}-i\omega_{1}}\delta_{\alpha_{2}^{\prime},\alpha_{1}}\,2\pi\delta_{i\omega_{1}^{\prime}-i\omega_{2}}\delta_{\alpha_{1}^{\prime},\alpha_{2}}\,A_{\infty}\left(\mathbf{k}_{1}^{\prime}\sigma_{1}^{\prime},\mathbf{k}_{2}^{\prime}\sigma_{2}^{\prime};\mathbf{k}_{2}\sigma_{2},\mathbf{k}_{1}\sigma_{1}\right),\nonumber 
\end{eqnarray}
where we defined 
\begin{equation}
A_{\infty}\left(\mathbf{k}_{1}^{\prime}\sigma_{1}^{\prime},\mathbf{k}_{2}^{\prime}\sigma_{2}^{\prime};\mathbf{k}_{1}\sigma_{1},\mathbf{k}_{2}\sigma_{2}\right)\equiv-2\pi\delta_{\mathbf{k}_{1}^{\prime}+\mathbf{k}_{2}^{\prime}-\mathbf{k}_{1}-\mathbf{k}_{2}}\,K\left(\hbar v\right)^{2}e^{-\frac{1}{2}\xi^{2}\left|\mathbf{k}_{1}^{\prime}-\mathbf{k}_{1}\right|^{2}}\delta_{\sigma_{1}^{\prime}\sigma_{1}}\delta_{\sigma_{2}^{\prime}\sigma_{2}},\label{eq:bare A}
\end{equation}
symmetric under the simultaneous exchange $\mathbf{k}_{1}^{\prime}\sigma_{1}^{\prime}\leftrightarrow\mathbf{k}_{2}^{\prime}\sigma_{2}^{\prime}$
and $\mathbf{k}_{1}\sigma_{1}\leftrightarrow\mathbf{k}_{2}\sigma_{2}$.

It is easy to see from the vertex flow equations (\ref{eq:D_Gamma_general})
that the locking of the replica and frequency indices, as present
in the bare vertex (\ref{eq:bare-vertex}), is preserved in the flow
(since the Green functions are frequency- and replica-diagonal). This
means the flowing vertex is always of the form $\Gamma_{\Lambda}(1_{\alpha_{1}i\omega_{1}}^{\prime},2_{\alpha_{2}i\omega_{2}}^{\prime};1_{\alpha_{1}i\omega_{1}},2_{\alpha_{2}i\omega_{2}})$
or $\Gamma_{\Lambda}(1_{\alpha_{1}i\omega_{1}}^{\prime},2_{\alpha_{2}i\omega_{2}}^{\prime};1_{\alpha_{2}i\omega_{2}},2_{\alpha_{1}i\omega_{1}})$.
Hence, in analogy to the bare vertex, we can write 
\begin{eqnarray}
\Gamma_{\Lambda}\left(1_{\alpha_{1}^{\prime}i\omega_{1}^{\prime}}^{\prime},2_{\alpha_{2}^{\prime}i\omega_{2}^{\prime}}^{\prime};1_{\alpha_{1}i\omega_{1}},2_{\alpha_{2}i\omega_{2}}\right) & = & 2\pi\delta_{i\omega_{1}^{\prime}-i\omega_{1}}2\pi\delta_{i\omega_{2}^{\prime}-i\omega_{2}}\delta_{\alpha_{1}^{\prime},\alpha_{1}}\delta_{\alpha_{2}^{\prime},\alpha_{2}}A_{\Lambda}^{\alpha_{1}^{\prime}i\omega_{1}^{\prime},\alpha_{2}^{\prime}i\omega_{2}^{\prime}}\left(\mathbf{k}_{1}^{\prime}\sigma_{1}^{\prime},\mathbf{k}_{2}^{\prime}\sigma_{2}^{\prime};\mathbf{k}_{1}\sigma_{1},\mathbf{k}_{2}\sigma_{2}\right)\label{eq:vertexForm}\\
 & - & 2\pi\delta_{i\omega_{2}^{\prime}-i\omega_{1}}2\pi\delta_{i\omega_{1}^{\prime}-i\omega_{2}}\delta_{\alpha_{2}^{\prime},\alpha_{1}}\delta_{\alpha_{1}^{\prime},\alpha_{2}}A_{\Lambda}^{\alpha_{1}^{\prime}i\omega_{1}^{\prime},\alpha_{2}^{\prime}i\omega_{2}^{\prime}}\left(\mathbf{k}_{1}^{\prime}\sigma_{1}^{\prime},\mathbf{k}_{2}^{\prime}\sigma_{2}^{\prime};\mathbf{k}_{2}\sigma_{2},\mathbf{k}_{1}\sigma_{1}\right),\nonumber 
\end{eqnarray}
with $A_{\Lambda}^{\alpha_{1}^{\prime}i\omega_{1}^{\prime},\alpha_{2}^{\prime}i\omega_{2}^{\prime}}\left(\mathbf{k}_{1}^{\prime}\sigma_{1}^{\prime},\mathbf{k}_{2}^{\prime}\sigma_{2}^{\prime};\mathbf{k}_{1}\sigma_{1},\mathbf{k}_{2}\sigma_{2}\right)$
symmetric under the simultaneous exchange $\alpha_{1}^{\prime}i\omega_{1}^{\prime}\leftrightarrow\alpha_{2}^{\prime}i\omega_{2}^{\prime}$
as well as $\mathbf{k}_{1}^{\prime}\sigma_{1}^{\prime}\leftrightarrow\mathbf{k}_{2}^{\prime}\sigma_{2}^{\prime}$
and $\mathbf{k}_{1}\sigma_{1}\leftrightarrow\mathbf{k}_{2}\sigma_{2}$
like $A_{\infty}$. 

Next, we need to leave out all terms on the rhs of Eqs. (\ref{eq:D_Sigma_general})
and (\ref{eq:D_Gamma_general}) where the sums $\sum_{\alpha_{3},\alpha_{4}}$
on the rhs provide an extra factor of $R$ as these vanish in the
replica limit, $\underset{R\rightarrow0}{\mathrm{lim}}\frac{1}{R}\sum_{\alpha=1}^{R}\left\langle \mathcal{O}\left(\bar{\psi}^{\alpha},\psi^{\alpha}\right)\right\rangle _{\psi}\propto\underset{R\rightarrow0}{\mathrm{lim}}\frac{1}{R}\sum_{\alpha=1}^{R}R\propto\underset{R\rightarrow0}{\mathrm{lim}}R=0$.
Note that fixing $1_{\alpha_{2}i\omega_{2}},2_{\alpha_{1}i\omega_{1}}$
in the first line of Eq. (\ref{eq:D_Gamma_general}) we can associate
$\alpha_{1}$ with the replica index from multi-index $3$ \emph{or}
$4$, in the second line we have no such choice and the third line
is always $\propto R$ and vanishes. If we would draw diagrams to
represent Eq. (\ref{eq:D_Gamma_general}), the replica limit condition
is equivalent of leaving out diagrams with internal fermion loops.
We find
\begin{eqnarray}
 &  & \partial_{\Lambda}\Gamma_{\Lambda}\left(1_{\alpha_{1}i\omega_{1}}^{\prime},2_{\alpha_{2}i\omega_{2}}^{\prime};1_{\alpha_{1}i\omega_{1}},2_{\alpha_{2}i\omega_{2}}\right)=\int_{3,3^{\prime},4,4^{\prime}}\label{eq:vertex flow}\\
 &  & \Gamma_{\Lambda}\left(1_{\alpha_{1}i\omega_{1}}^{\prime},2_{\alpha_{2}i\omega_{2}}^{\prime};3_{\alpha_{1}i\omega_{1}},4_{\alpha_{2}i\omega_{2}}\right)\left(\left[G_{\Lambda}(\alpha_{1}i\omega_{1})\right]_{3,3^{\prime}}\left[\dot{G}_{\Lambda}(\alpha_{2}i\omega_{2})\right]_{4,4^{\prime}}+\dot{G}\leftrightarrow G\right)\Gamma_{\Lambda}\left(3_{\alpha_{1}i\omega_{1}}^{\prime},4_{\alpha_{2}i\omega_{2}}^{\prime};1_{\alpha_{1}i\omega_{1}},2_{\alpha_{2}i\omega_{2}}\right)\nonumber \\
 & + & \Gamma_{\Lambda}\left(1_{\alpha_{1}i\omega_{1}}^{\prime},4_{\alpha_{2}i\omega_{2}}^{\prime};3_{\alpha_{1}i\omega_{1}},2_{\alpha_{2}i\omega_{2}}\right)\left(\left[G_{\Lambda}(\alpha_{1}i\omega_{1})\right]_{3,3^{\prime}}\left[\dot{G}_{\Lambda}(\alpha_{2}i\omega_{2})\right]_{4,4^{\prime}}+\dot{G}\leftrightarrow G\right)\Gamma_{\Lambda}\left(3_{\alpha_{1}i\omega_{1}}^{\prime},2_{\alpha_{2}i\omega_{2}}^{\prime};1_{\alpha_{1}i\omega_{1}},4_{\alpha_{2}i\omega_{2}}\right),\nonumber 
\end{eqnarray}

\paragraph{Self-energy and vertex flow.\textemdash{}}

Eventually, for the DOS we are interested in the Green function which
involves the self-energy. Employing the replica-frequency locking
of the vertex for the self-energy flow Eq. (\ref{eq:D_Sigma_general}),
we find
\begin{align*}
\partial_{\Lambda}\Sigma_{\Lambda}\left(\alpha_{1}i\omega_{1}\mathbf{k}_{1}\right)_{\sigma_{1}^{\prime},\sigma_{1}} & =-\int_{22^{\prime}\alpha_{2},i\omega_{2}}\left[\dot{G}_{\Lambda}(\alpha_{2}i\omega_{2})\right]_{2,2^{\prime}}\Gamma_{\Lambda}\left(1_{\alpha_{1}i\omega_{1}}^{\prime},2_{\alpha_{2}i\omega_{2}}^{\prime};1_{\alpha_{1}i\omega_{1}},2_{\alpha_{2}i\omega_{2}}\right).
\end{align*}
Applying Eq. (\ref{eq:vertexForm}), we find that only the second
part avoids the replica sum leading to $\propto R$. The Green function
locks \emph{all} frequencies and replica indices appearing on the
rhs of the self-energy flow equation,
\begin{equation}
\partial_{\Lambda}\Sigma_{\Lambda}\left(\alpha i\omega\mathbf{k}_{1}\right)_{\sigma_{1}^{\prime},\sigma_{1}}=\sum_{\sigma_{2},\sigma_{2}^{\prime}}\int_{\mathbf{k}_{2}}\left[\dot{G}_{\Lambda}\left(\alpha i\omega\mathbf{k}_{2}\right)\right]_{\sigma_{2},\sigma_{2}^{\prime}}A_{\Lambda}^{\alpha i\omega,\alpha i\omega}\left(\mathbf{k}_{1}\sigma_{1}^{\prime},\mathbf{k}_{2}\sigma_{2}^{\prime};\mathbf{k}_{2}\sigma_{2},\mathbf{k}_{1}\sigma_{1}\right).\label{eq:D_Sigma}
\end{equation}

In Eq. (\ref{eq:D_Sigma}), the function $A$ only appears with equal
replica and frequency indices. We insert this structure in Eq. (\ref{eq:vertex flow})
and obtain 
\begin{eqnarray}
 &  & \partial_{\Lambda}A_{\Lambda}^{\alpha i\omega,\alpha i\omega}\left(\mathbf{k}_{1}^{\prime}\sigma_{1}^{\prime},\mathbf{k}_{2}^{\prime}\sigma_{2}^{\prime};\mathbf{k}_{1}\sigma_{1},\mathbf{k}_{2}\sigma_{2}\right)=\int_{\mathbf{k}_{3},\mathbf{k}_{4}}\sum_{\sigma_{3}^{\prime}\sigma_{3},\sigma_{4}^{\prime}\sigma_{4}}\label{eq:D_A}\\
 &  & A_{\Lambda}^{\alpha i\omega,\alpha i\omega}\left(\mathbf{k}_{1}^{\prime}\sigma_{1}^{\prime},\mathbf{k}_{2}^{\prime}\sigma_{2}^{\prime};\mathbf{k}_{3}\sigma_{3},\mathbf{k}_{4}\sigma_{4}\right)\left(\left[G_{\Lambda}\left(\alpha i\omega\mathbf{k}_{3}\right)\right]_{\sigma_{3},\sigma_{3}^{\prime}}\left[\dot{G}_{\Lambda}\left(\alpha i\omega\mathbf{k}_{4}\right)\right]_{\sigma_{4},\sigma_{4}^{\prime}}+\dot{G}\leftrightarrow G\right)A_{\Lambda}^{\alpha i\omega,\alpha i\omega}\left(\mathbf{k}_{3}\sigma_{3}^{\prime},\mathbf{k}_{4}\sigma_{4}^{\prime};\mathbf{k}_{1}\sigma_{1},\mathbf{k}_{2}\sigma_{2}\right)\nonumber \\
 & + & A_{\Lambda}^{\alpha i\omega,\alpha i\omega}\left(\mathbf{k}_{1}^{\prime}\sigma_{1}^{\prime},\mathbf{k}_{4}\sigma_{4}^{\prime};\mathbf{k}_{3}\sigma_{3},\mathbf{k}_{2}\sigma_{2}\right)\left(\left[G_{\Lambda}\left(\alpha i\omega\mathbf{k}_{3}\right)\right]_{\sigma_{3},\sigma_{3}^{\prime}}\left[\dot{G}_{\Lambda}\left(\alpha i\omega\mathbf{k}_{4}\right)\right]_{\sigma_{4},\sigma_{4}^{\prime}}+\dot{G}\leftrightarrow G\right)A_{\Lambda}^{\alpha i\omega,\alpha i\omega}\left(\mathbf{k}_{3}\sigma_{3}^{\prime},\mathbf{k}_{2}^{\prime}\sigma_{2}^{\prime};\mathbf{k}_{1}\sigma_{1},\mathbf{k}_{4}\sigma_{4}\right).\nonumber 
\end{eqnarray}
We can now drop the replica index $\alpha$ from our intermediate
flow equations (\ref{eq:D_Sigma}) and (\ref{eq:D_A}) and proceed
to specify the flow parameter $\Lambda$ which was general so far.

\paragraph{Matsubara frequency cutoff.\textemdash{}}

In its standard application to systems with inelastic (true) interactions,
the fRG flow equations contain frequency integrals on the rhs \cite{Metzner2012}.
This integral is absent in Eq. (\ref{eq:D_A}) due to the elastic
structure of the disorder induced interaction vertex. We can take
this to our advantage and choose a Matsubara cutoff scheme which will
allow exact integration of the flow equations. In the Matsubara cutoff
scheme a multiplicative cutoff to the bare Green function is employed
$G_{0,\Lambda}(1_{i\omega_{1}})=\theta\left(|i\omega_{1}|-\Lambda\right)G_{0}(1_{i\omega_{1}})$,
the corresponding single scale propagator reads $\dot{G}_{\Lambda}(1_{i\omega_{1}})=\delta\left(|i\omega_{1}|-\Lambda\right)\tilde{G}_{\Lambda}(1_{i\omega_{1}})$
and $\dot{G}_{\Lambda}(1_{i\omega_{1}})G_{\Lambda}(1_{i\omega_{2}})=\delta\left(|i\omega_{1}|-\Lambda\right)\Theta\left(|i\omega_{2}|-\Lambda\right)\tilde{G}_{\Lambda}(1_{i\omega_{1}})\tilde{G}_{\Lambda}(1_{i\omega_{2}})$
where $\tilde{G}_{\Lambda}(1_{i\omega_{1}})=\left[G_{0}^{-1}(1_{i\omega_{1}})-\Sigma_{\Lambda}(1_{i\omega_{1}})\right]^{-1}$
\cite{Metzner2012} and $\theta(0)=1/2$ is understood by Morris Lemma
\cite{Morris1993}.

We find
\begin{eqnarray*}
 &  & \partial_{\Lambda}\Sigma_{\Lambda}\left(i\omega\mathbf{k}_{1}\right)_{\sigma_{1}^{\prime},\sigma_{1}}\:=\:\delta\left(|i\omega|-\Lambda\right)\sum_{\sigma_{2},\sigma_{2}^{\prime}}\int_{\mathbf{k}_{2}}\tilde{G}_{\Lambda}\left(i\omega,\mathbf{k}_{2}\right)_{\sigma_{2},\sigma_{2}^{\prime}}A_{\Lambda}^{i\omega,i\omega}\left(\mathbf{k}_{1}\sigma_{1}^{\prime},\mathbf{k}_{2}\sigma_{2}^{\prime};\mathbf{k}_{2}\sigma_{2},\mathbf{k}_{1}\sigma_{1}\right),\\
 &  & \partial_{\Lambda}A_{\Lambda}^{i\omega,i\omega}\left(\mathbf{k}_{1}^{\prime}\sigma_{1}^{\prime},\mathbf{k}_{2}^{\prime}\sigma_{2}^{\prime};\mathbf{k}_{1}\sigma_{1},\mathbf{k}_{2}\sigma_{2}\right)\:=\:2\delta\left(|i\omega|-\Lambda\right)\Theta\left(|i\omega|-\Lambda\right)\int_{\mathbf{k}_{3},\mathbf{k}_{4}}\sum_{\sigma_{3}^{\prime}\sigma_{3},\sigma_{4}^{\prime}\sigma_{4}}\\
 &  & A_{\Lambda}^{i\omega,i\omega}\left(\mathbf{k}_{1}^{\prime}\sigma_{1}^{\prime},\mathbf{k}_{2}^{\prime}\sigma_{2}^{\prime};\mathbf{k}_{3}\sigma_{3},\mathbf{k}_{4}\sigma_{4}\right)\left(\left[\tilde{G}_{\Lambda}\left(i\omega\mathbf{k}_{3}\right)\right]_{\sigma_{3},\sigma_{3}^{\prime}}\left[\tilde{G}_{\Lambda}\left(i\omega\mathbf{k}_{4}\right)\right]_{\sigma_{4},\sigma_{4}^{\prime}}\right)A_{\Lambda}^{i\omega,i\omega}\left(\mathbf{k}_{3}\sigma_{3}^{\prime},\mathbf{k}_{4}\sigma_{4}^{\prime};\mathbf{k}_{1}\sigma_{1},\mathbf{k}_{2}\sigma_{2}\right)\\
 & + & A_{\Lambda}^{i\omega,i\omega}\left(\mathbf{k}_{1}^{\prime}\sigma_{1}^{\prime},\mathbf{k}_{4}\sigma_{4}^{\prime};\mathbf{k}_{3}\sigma_{3},\mathbf{k}_{2}\sigma_{2}\right)\left(\left[\tilde{G}_{\Lambda}\left(i\omega\mathbf{k}_{3}\right)\right]_{\sigma_{3},\sigma_{3}^{\prime}}\left[\tilde{G}_{\Lambda}\left(i\omega\mathbf{k}_{4}\right)\right]_{\sigma_{4},\sigma_{4}^{\prime}}\right)A_{\Lambda}^{i\omega,i\omega}\left(\mathbf{k}_{3}\sigma_{3}^{\prime},\mathbf{k}_{2}^{\prime}\sigma_{2}^{\prime};\mathbf{k}_{1}\sigma_{1},\mathbf{k}_{4}\sigma_{4}\right),
\end{eqnarray*}
Assuming $|\omega|>0$, we can now integrate both flow equations exactly
over $\Lambda$ from $\Lambda=\infty$ to $\Lambda=0$ to find the
physical self-energy $\Sigma=\Sigma_{\Lambda=0}$ and vertex function
$A=A_{\Lambda=0}$. The initial condition for the interaction vertex
is the bare interaction. Writing simply $G$ instead of $\tilde{G}_{\Lambda=0}$,
we find 
\begin{eqnarray}
 &  & \Sigma\left(i\omega\mathbf{k}_{1}\right)_{\sigma_{1}^{\prime},\sigma_{1}}\:=\:-\sum_{\sigma_{2},\sigma_{2}^{\prime}}\int_{\mathbf{k}_{2}}G\left(i\omega\mathbf{k}_{2}\right)_{\sigma_{2},\sigma_{2}^{\prime}}A^{i\omega,i\omega}\left(\mathbf{k}_{1}\sigma_{1}^{\prime},\mathbf{k}_{2}\sigma_{2}^{\prime};\mathbf{k}_{2}\sigma_{2},\mathbf{k}_{1}\sigma_{1}\right),\label{eq:S_self-consistency}\\
 &  & A^{i\omega,i\omega}\left(\mathbf{k}_{1}^{\prime}\sigma_{1}^{\prime},\mathbf{k}_{2}^{\prime}\sigma_{2}^{\prime};\mathbf{k}_{1}\sigma_{1},\mathbf{k}_{2}\sigma_{2}\right)\:=\:A_{\infty}\left(\mathbf{k}_{1}^{\prime}\sigma_{1}^{\prime},\mathbf{k}_{2}^{\prime}\sigma_{2}^{\prime};\mathbf{k}_{1}\sigma_{1},\mathbf{k}_{2}\sigma_{2}\right)-\int_{\mathbf{k}_{3},\mathbf{k}_{4}}\sum_{\sigma_{3}^{\prime}\sigma_{3},\sigma_{4}^{\prime}\sigma_{4}}\label{eq:D_self-consistency}\\
 &  & \:\:A^{i\omega,i\omega}\left(\mathbf{k}_{1}^{\prime}\sigma_{1}^{\prime},\mathbf{k}_{2}^{\prime}\sigma_{2}^{\prime};\mathbf{k}_{3}\sigma_{3},\mathbf{k}_{4}\sigma_{4}\right)\left(\left[G\left(i\omega\mathbf{k}_{3}\right)\right]_{\sigma_{3},\sigma_{3}^{\prime}}\left[G\left(i\omega\mathbf{k}_{4}\right)\right]_{\sigma_{4},\sigma_{4}^{\prime}}\right)A^{i\omega,i\omega}\left(\mathbf{k}_{3}\sigma_{3}^{\prime},\mathbf{k}_{4}\sigma_{4}^{\prime};\mathbf{k}_{1}\sigma_{1},\mathbf{k}_{2}\sigma_{2}\right)\nonumber \\
 &  & +A^{i\omega,i\omega}\left(\mathbf{k}_{1}^{\prime}\sigma_{1}^{\prime},\mathbf{k}_{4}\sigma_{4}^{\prime};\mathbf{k}_{3}\sigma_{3},\mathbf{k}_{2}\sigma_{2}\right)\left(\left[G\left(i\omega\mathbf{k}_{3}\right)\right]_{\sigma_{3},\sigma_{3}^{\prime}}\left[G\left(i\omega\mathbf{k}_{4}\right)\right]_{\sigma_{4},\sigma_{4}^{\prime}}\right)A^{i\omega,i\omega}\left(\mathbf{k}_{3}\sigma_{3}^{\prime},\mathbf{k}_{2}^{\prime}\sigma_{2}^{\prime};\mathbf{k}_{1}\sigma_{1},\mathbf{k}_{4}\sigma_{4}\right).\nonumber 
\end{eqnarray}
Instead of the usual coupled fRG \emph{flow} equations that have to
be integrated, we thus have rephrased the disorder problem in terms
of the coupled self-consistent Eqns. (\ref{eq:S_self-consistency})
and (\ref{eq:D_self-consistency}). Note that the above derivation
did not depend on the three (or $N$-) particle vertex truncation
in Eq. (\ref{eq:D_Gamma_general}) and thus, an extended set of coupled
self-consistency equations would still be exact.

We turn back to our initial goal to find the DOS at the nodal point
$E=0$. For this, we need the retarded real frequency self-energy,
see Eq. (\ref{eq:DOS}) in the main text, that is connected to $\Sigma\left(i\omega\right)$
by an analytical continuation $i\omega=0+i0^{+}$
where $0^{+}$ is a positive real infinitesimal. After this step,
we drop the frequency variable from now on. Let us emphasize that
the appearance of a single frequency in the hierarchy of self-consistent
equations is a remnant of the elastic nature of disorder scattering.

\paragraph{Solution correct to order $K^{2}$.\textemdash{}}

Even the set of self-consistency equations (\ref{eq:S_self-consistency})
and (\ref{eq:D_self-consistency}) (with the three-particle vertex
dropped) is difficult to solve without further approximations. To
obtain the self-energy correct to at least $\mathcal{O}(K^{2})$,
on the rhs of Eq. (\ref{eq:D_self-consistency}), is is sufficient
to use the bare vertex Eq. (\ref{eq:bare A}). This is a pragmatic
approach, which, however still goes beyond existing studies in the
literature. We obtain from Eqns. (\ref{eq:bare A}) and (\ref{eq:D_self-consistency})
\begin{eqnarray*}
 &  & A\left(\mathbf{k}_{1}^{\prime}\sigma_{1}^{\prime},\mathbf{k}_{2}^{\prime}\sigma_{2}^{\prime};\mathbf{k}_{1}\sigma_{1},\mathbf{k}_{2}\sigma_{2}\right)=A_{\infty}\left(\mathbf{k}_{1}^{\prime}\sigma_{1}^{\prime},\mathbf{k}_{2}^{\prime}\sigma_{2}^{\prime};\mathbf{k}_{1}\sigma_{1},\mathbf{k}_{2}\sigma_{2}\right)-K^{2}\left(\hbar v\right)^{4}\int_{\mathbf{k}_{3}}\\
 & \times & e^{-\frac{1}{2}\xi^{2}\left|\mathbf{k}_{1}^{\prime}-\mathbf{k}_{3}\right|^{2}}\left[G\left(\mathbf{k}_{3}\right)\right]_{\sigma_{1}^{\prime},\sigma_{1}}\left(\left[G\left(\mathbf{k}_{1}^{\prime}+\mathbf{k}_{2}^{\prime}-\mathbf{k}_{3}\right)\right]_{\sigma_{2}^{\prime},\sigma_{2}}+\left[G\left(\mathbf{k}_{3}+\mathbf{k}_{2}-\mathbf{k}_{1}^{\prime}\right)\right]_{\sigma_{2}^{\prime}\sigma_{2}}\right)e^{-\frac{1}{2}\xi^{2}\left|\mathbf{k}_{1}-\mathbf{k}_{3}\right|^{2}},
\end{eqnarray*}
and further specialize to the spin-momentum structure needed for the
self-energy flow Eq. (\ref{eq:D_Sigma})
\begin{eqnarray}
 &  & A\left(\mathbf{k}_{1}\sigma_{1}^{\prime},\mathbf{k}_{2}\sigma_{2}^{\prime};\mathbf{k}_{2}\sigma_{2},\mathbf{k}_{1}\sigma_{1}\right)=-K\left(\hbar v\right)^{2}e^{-\frac{1}{2}\xi^{2}\left|\mathbf{k}_{2}-\mathbf{k}_{1}\right|^{2}}\delta_{\sigma_{1}^{\prime}\sigma_{1}}\delta_{\sigma_{2}^{\prime}\sigma_{2}}-K^{2}\left(\hbar v\right)^{4}\int_{\mathbf{k}_{3}}\label{eq:D_A_O(K2)}\\
 & \times & e^{-\frac{1}{2}\xi^{2}\left|\mathbf{k}_{1}-\mathbf{k}_{3}\right|^{2}}\left[G\left(\mathbf{k}_{3}\right)\right]_{\sigma_{1}^{\prime},\sigma_{1}}\left(\left[G\left(\mathbf{k}_{1}^{\prime}+\mathbf{k}_{2}^{\prime}-\mathbf{k}_{3}\right)\right]_{\sigma_{2}^{\prime},\sigma_{2}}+\left[G\left(\mathbf{k}_{3}+\mathbf{k}_{2}-\mathbf{k}_{1}^{\prime}\right)\right]_{\sigma_{2}^{\prime}\sigma_{2}}\right)e^{-\frac{1}{2}\xi^{2}\left|\mathbf{k}_{2}-\mathbf{k}_{3}\right|^{2}}.\nonumber 
\end{eqnarray}

We combine Eq. (\ref{eq:D_A_O(K2)}) with (\ref{eq:S_self-consistency})
and find the final self-consistency equation. Relabeling $\mathbf{k}_{1}\rightarrow\mathbf{k}$,
$\mathbf{k}_{2}\rightarrow\mathbf{q}$ and $\mathbf{k}_{3}\rightarrow\mathbf{p}$
and using $``\cdot$'' to indicate matrix products for the 2x2 matrix-valued
Green functions, we arrive at Eq. (\ref{eq:self-consistency_short})
from the main text: 
\begin{align}
\Sigma\left(\mathbf{k}\right) & =K\left(\hbar v\right)^{2}\int_{\mathbf{q}}e^{-\frac{1}{2}\xi^{2}\left|\mathbf{q}-\mathbf{k}\right|^{2}}\,G(\mathbf{q})\label{eq:self-consistency}\\
 & +K^{2}\left(\hbar v\right)^{4}\int_{\mathbf{q},\mathbf{p}}e^{-\frac{1}{2}\xi^{2}(\left|\mathbf{k}-\mathbf{p}\right|^{2}+\left|\mathbf{q}-\mathbf{p}\right|^{2})}\,G(\mathbf{p})\cdot G(\mathbf{q})\cdot\left[G(\mathbf{k}+\mathbf{q}-\mathbf{p})+G(\mathbf{p})\right].\nonumber 
\end{align}
If the feedback of the flowing vertex $A$ to the rhs of its own
flow equation would be considered, this would yield two equations
for $\Sigma$ and $A$ to be solved self-consistently. 

\paragraph{Numerical solution of self-consistency equations.\textemdash{}}

The self-consistency equation (\ref{eq:self-consistency}) can be
solved numerically by iteration. We use dimensionless units (measuring
momenta in $1/\xi$ and energies in $\hbar v/\xi$) and the dimensionless
self-energy in $d=2$ (at the nodal point) can be parametrized as
\begin{equation}
\frac{\Sigma_{d=2}\left(\mathbf{x}=\mathbf{k}\xi\right)}{\hbar v/\xi}=m_{2}\left(x\right)\left\{ \sigma_{x}\cos\left[\phi\right]+\sigma_{y}\sin\left[\phi\right]\right\} +iM_{2}\left(x\right),
\end{equation}
with $x,\phi$ polar coordinates. The term $M_{2}\left(x\right)$
has to be purely real (to avoid a spontaneous creation of chemical
potential) and $>0$ for the retarded self energy. As a result, on
the rhs of Eq. (\ref{eq:self-consistency}), we can chose $\mathbf{k}$
in say, the x-direction and also take only the $\sigma_{x}$ component
of the product of Green functions (it can be checked that all other
components vanish). The final self-consistency loop is then only for
the functions $m_{2}\left(x\right)$ and $M_{2}(x)$, which turn out
to be rather smooth. They can be discretized on a geometric grid for
the variable $x$, the angular integrations can be done using a linearly
spaced integration grid for the angles. We made sure that our results
are converged with respect to the resolution of the discretization
grids. Once $m_{2},\,M_{2}$ do not change any more under insertion
on the rhs of Eq. (\ref{eq:self-consistency}), the DOS is computed
from Eq. (\ref{eq:DOS}) using interpolation of the integrand and
quadrature integration. Likewise, in $d=3$, the same strategy is
applied using a parametrization in spherical coordinates $x,\phi,\theta$:
\begin{equation}
\frac{\Sigma_{d=3}\left(\mathbf{x}=\mathbf{k}\xi\right)}{\hbar v/\xi}=m_{3}\left(x\right)\left(\sin\left[\theta\right]\left\{ \sigma_{x}\cos\left[\phi\right]+\sigma_{y}\sin\left[\phi\right]\right\} +\sigma_{z}\cos\left[\theta\right]\right)+iM_{3}\left(x\right).
\end{equation}

\enddocument